\documentclass[12pt]{article}  
\usepackage{epsfig}  
 1 
\def\as{\alpha_{\rm s}} 
\def\aem{\alpha_{\rm em}} 
 
\def\dd{\partial}

\def\percent{{\%\ }} 
 
\newcommand{\half}{\mbox{\small $\frac{1}{2}$}}

\def\gev{{\,\mbox{GeV}  }} 
\def\gevs{{\,\mbox{GeV}^2}} 
\def\gevms{{\,\mbox{GeV}^{-2}}}

\def\({\left(} 
\def\){\right)}

\def\rt{r_{\perp}} 
\def\rhot{\rho_{\perp}} 
 
\def\kt{k_{\perp}}

\def\eg{\hbox{\it e.g.}}

\def\ket#1{\left| #1\right\rangle}

\def\beq{\begin{equation}} 
\def\eeq{\end{equation}} 
\def\bea{\begin{eqnarray}} 
\def\eea{\end{eqnarray}} 
\def\eq#1{{\mbox{Eq.\hspace{1mm}(\ref{#1})}}} 
\def\eqs#1#2{{\mbox{Eqs.\hspace{1mm}(\ref{#1})--(\ref{#2})}}} 
\def\fig#1{{\mbox{Fig.\hspace{1mm}\ref{#1}}}}

\def\scrbox#1{\mbox{\scriptsize #1}}

\def\npb#1#2#3{    {\it Nucl.\ Phys.\ }{\bf B#1} (#2) #3} 
\def\npa#1#2#3{    {\it Nucl.\ Phys.\ }{\bf A#1} (#2) #3} 
\def\plb#1#2#3{    {\it Phys.\ Lett.\ }{\bf B#1} (#2) #3} 
\def\prd#1#2#3{    {\it Phys.\ Rev.\ }{\bf D#1} (#2) #3} 
\def\prep#1#2#3{   {\it Phys.\ Rep.\ }{\bf #1} (#2) #3}

\def\zpc#1#2#3{    {\it Z.\ Phys.\ }{\bf C#1} (#2) #3}

\def\epj#1#2#3{    {\it Eur.\ Phys.\ J.\ }       {\bf #1} (#2) #3}

\newcommand{\email}[1]{${\!}^{\scrbox{#1)}}$}

\newcommand{\sigin}{\sigma_{\scrbox{input}}} 
\newcommand{\xgdglap}{xG^{\scrbox{DGLAP}}}

\newcommand{\chisquare}{\chi^2/\mbox{n.d.f.}} 
\newcommand{\jpsi}{J/\psi} 
 
\newcommand{\sd}{\sigma_{\mbox{\scriptsize dipole}}} 
\newcommand{\dsize}[1]{\mathbf{x}_{#1}} 
\newcommand{\sdsize}[1]{x^2_{#1}} 
\newcommand{\cA}{{\cal{A}}} 
\newcommand{\tn}{\tilde N} 
\def\df2dlnq2{\dd{F_2}/\dd\log{Q^2}}

\relax 
\begin{document} 
\begin{titlepage} 
\noindent 
\begin{flushright} 
\parbox[t]{14em}{
\begin{tabular}{ll}
TAUP &   2721/2003\\ 
DESY &   03-011\\
hep-ph/ & 0302010
\end{tabular}
} 
\end{flushright} 
\vspace{1cm} 
\begin{center} 
{\Large \bf 
    ${\mathbf J/\psi}$ Photo- and DIS Production via Nonlinear Evolution}
  \\[4ex] 
\begin{center}\large{ 
        E.~Gotsman  $^{a}$ \email{1},  
        E.~Levin    $^{a,\,b}$ \email{2},  
        M.~Lublinsky $^{b}$  \email{3},  
        U.~Maor     $^{a}$ \email{4} and  
        E.~Naftali  $^{a}$ \email{5}} 
\end{center} 
 
\footnotetext{\email{1} gotsman@post.tau.ac.il } 
\footnotetext{\email{2} leving@post.tau.ac.il } 
\footnotetext{\email{3} lublinm@mail.desy.de } 
\footnotetext{\email{4} maor@post.tau.ac.il } 
\footnotetext{\email{5} erann@post.tau.ac.il } 
\vfill 
{\it  $^{a)}$ School of Physics and Astronomy}\\ 
{\it  Raymond and Beverly Sackler Faculty of Exact Science}\\ 
{\it  Tel Aviv University, Tel Aviv, 69978, ISRAEL}\\[4.5ex] 
\vfill 
 {\it $^{b)}$ DESY Theory Group}\\  
{\it 22603, Hamburg, GERMANY}\\[4.5ex]

\end{center} 
~\,\, 
\vspace{1cm}

{\samepage {\large \bf Abstract:}}  
\vfill  
 We apply our solution of the nonlinear evolution equation to the case of
 $J/\psi$ photo and DIS production on nucleons and nuclei targets.  The
 uncertainty in the $\jpsi$ wave function normalization due to Fermi motion
 is treated as a free parameter. We obtain good reproduction of the HERA
 experimental data on a proton target.  Calculations of $\jpsi$ mesons
 coherent production on nuclei targets are presented and discussed.  Our
 analysis supports the conclusions reached in our previous studies, stressing
 the importance of nonlinear evolution in the kinematical domain of
 high-density QCD.

\vfill 
 
\end{titlepage} 
 \section{Introduction} \label{sec:introduction} 
 It was recognized long ago \cite{GLR} that the interactions of virtual
 photons at high-energies and low virtualities are governed by nonlinear
 evolution. Over the years, different theoretical methods of calculating
 processes in the kinematical region of high-density QCD (hdQCD) have been
 proposed \cite{MU86,MV,SAT,ELTHEORY,BA,KO}, leading to the construction of
 an equation, which incorporates both linear evolution due to parton
 splitting, and nonlinear evolution due to the recombination of partons at
 high-density.  This equation is given in \eq{eq:BK} below, using the
 notations of \cite{KO}, for $\tn(\rt,x; b)$, the imaginary part of the
 amplitude of a dipole of size $\rt$, which scatters elastically at impact
 parameter $b$.
 
In this paper we extend our recent investigations \cite{NLE01,NLE02} of 
 the nonlinear evolution to vector mesons, focusing on $\jpsi$ photo 
 \cite{ZEUSpsiPhoto2002} and DIS \cite{ZEUSpsiDIS2002} production.  As this 
 process is characterized by a scale of the order of the charmed quark mass, 
 $m_c$, it is considered a primary candidate for investigating the 
 kinematical region of QCD on the boundary between perturbative and hdQCD. 

In our previous publications \cite{GLMNpsi,GLMNpsi2} we demonstrated that the
 cross section experimental data for $\jpsi$ production is well reproduced if
 one employs two damping factors, calculated using a Glauber like
 shadowing-correction (SC) formalism \cite{SC}.  Broadly speaking, each
 damping factor may be considered as one iteration in an iterative procedure
 for determining $\tn(\rt,x; b)$.  These iterations correspond to Glauber
 rescatterings where the first iteration is attributed to SC due to the
 passage of a dipole through the target, and the second iteration is
 attributed to one small $x$ gluon emitted by the dipole prior to the
 interaction.  We believe that the SC formalism was, at the time, an
 important preliminary step towards the more rigorous treatment of
 \cite{NLE02}.

A particular outcome of \cite{GLMNpsi2}, is that the $\jpsi$ experimental
 data for the forward differential cross section slope, $B$, can be well
 reproduced by assuming that the profile function in the impact parameter
 space is the Fourier transform of an electromagnetic dipole-like form
 factor.  Our numerical calculations of $B$ produced a satisfactory fit to
 the experimental data with a hadron radius $R^2=10\,\gevms$.  On the other
 hand, in \cite{NLE02}, a good fit to the $F_2$ data was obtained with
 $R^2=3.1\,\gevms$ for an exponential form factor, and $R^2=4.5\,\gevms$, for
 a dipole-like form factor.  We consider the radii in \cite{NLE02} as
 effective radii which are reduced due to ({\it i}) the presence of inelastic
 diffractive processes; and ({\it ii}) the observation that in the simple
 expression typically used for $\sd$, the dipole-proton cross section, a
 small anomalous dimension has been assumed.  As we shall see, when
 calculating the cross section for $\jpsi$ production, it is necessary to
 include an effective $b$-dependence, which we choose to extract from the
 HERA experimental data.

This paper is organized as follows: in section \ref{sec:N} we briefly review
 the nonlinear equation and the approximate numerical solution to it, as
 obtained in \cite{NLE02}; in section \ref{sec:cross} we calculate the
 integrated cross section for $\jpsi$ photo- and DIS- production, and compare
 to the relevant experimental data; in section \ref{sec:nuc} we present our
 predictions for the production of $\jpsi$ from scattering of electrons on
 heavy nuclei. Our summary and conclusions are given in section
 \ref{sec:summary}.

\section{The Dipole Scattering Amplitude} \label{sec:N} 

The imaginary part of the dipole scattering amplitude, $\tn$, is a solution
 of a nonlinear evolution equation, which characterizes the low $x$ behavior
 of the parton densities, while taking into account hdQCD effects, thereby
 obeying the unitarity constraints.  The equation describes the interaction
 with a target of a parent dipole, of size $\dsize{01}$, and of two dipoles,
 of sizes $\dsize{12}$ and $\dsize{02}$, which were produced by the dipole of
 size $\dsize{01}$.  The probability for the decay of the dipole of size
 $\dsize{01}$ is given by the square of its wave function, which, in a
 simplified form, can be written as $\sdsize{01}/\sdsize{02}\sdsize{12}$.

Each of the produced dipoles can interact with the target independently, with
 respective amplitudes of $\tn(\dsize{12},y;{\mathbf{b}-\half\dsize{02}})$
 and $\tn(\dsize{02},y;{\mathbf{b}-\half\dsize{12}})$, where $y$ denotes the
 rapidity variable and $b$ the impact parameter.  However, adding these
 contributions clearly overestimates the dipole-nucleon interaction, since
 one must consider the probability that during the interaction, one dipole is
 in the shadow of the other. This negative correction factor is given by
 $-\tn(\dsize{12},y;\mathbf{b}-\half\dsize{02})\tilde
 N(\dsize{02},y;\mathbf{b}-\half\dsize{12})$.
 
Thus, $\tn$ can be written in the form  
\begin{eqnarray} 
\lefteqn{\tn({\mathbf{x_{01}}},Y;b) = \tilde 
N({\mathbf{x_{01}}},Y_0;b)\, {\rm exp}\left[-\frac{2 \,C_F\,\as}{\pi} 
\,\ln\left( \frac{{\mathbf{x^2_{01}}}}{\rho^2}\right)(Y-Y_0)\right ]\, + 
}\nonumber \\ & & \frac{C_F\,\as}{\pi^2}\,\int_{Y_0}^Y dy \, {\rm 
exp}\left[-\frac{2 \,C_F\,\as}{\pi} \,\ln\left( 
\frac{{\mathbf{x^2_{01}}}}{\rho^2}\right)(Y-y)\right ]\,\times 
\label{eq:BK} 
\\  
& & \int_{\rho} d^2 {\mathbf{x_{2}}}  
\frac{{\mathbf{x^2_{01}}}}{{\mathbf{x^2_{02}}}\, 
{\mathbf{x^2_{12}}}} \nonumber 
\left(\,2\,\tn({\mathbf{x_{02}}},y;{ \mathbf{ b- 
\frac{1}{2} x_{12}}}) 
-\tn({\mathbf{x_{02}}},y;{ \mathbf{ b - 
\frac{1}{2} 
x_{12}}})\tn({\mathbf{x_{12}}},y;{ \mathbf{ b- \frac{1}{2} 
x_{02}}})\right)\,, 
\end{eqnarray} 
 where, $Y=-\ln x$, $Y_0=-\ln x_0$ and $\rho$ is an ultraviolet cutoff, which
 does not appear in the physical quantities.

The linear part of (\ref{eq:BK}) is the LO BFKL equation \cite{BFKL}, which
 describes the evolution of the multiplicity of the fixed size color dipoles
 as a function of the energy $Y$. The equation sums the high twist
 contributions. Note, that the linear part of (\ref{eq:BK}) also has higher
 twist contributions and the main contribution of the nonlinear part is to
 the leading twist (see \cite{MU86} for general arguments and \cite{HTM} for
 explicit calculations).

For completeness, we provide a brief description of the steps taken in
 \cite{NLE02} for obtaining an approximate solution for
 $\tn({\mathbf{x_{01}}},Y;b)$.
 
The initial conditions for (\ref{eq:BK}) were taken at $x_0=10^{-2}$ in the
 eikonal approximation, accounting for multiple dipole-target interactions:
\begin{equation}\label{initialcondition} 
\tn(\dsize{01},x_0;b)\,=\,1\,-\,e^{-\frac{1}{2}\sigin(\dsize{01},x_0)\,S(b)}, 
\end{equation} 
where 
\begin{equation}\label{SBP} 
\sigin(\dsize{01},x_0) = 
\frac{\alpha_s\,\pi^2}{N_c}\sdsize{01}\xgdglap(x_0,4/\sdsize{01}), 
\end{equation} 
and $S(b)$ is the profile function in impact parameter space.  As stated, 
we found that the experimental data of the differential cross section slope, 
$B$, is well described by the following profile function, which is the 
Fourier transform of electromagnetic dipole-like form factor: 
\begin{equation}\label{dipS} 
S(b)=\frac{2}{\pi  
R^2}\frac{\sqrt{8}b}{R}K_1(\frac{\sqrt{8}b}{R}). 
\end{equation} 

As a first step, the $b$-dependence of (\ref{eq:BK}) was neglected. Thus, all
 twist contributions for the evolution were summed, using the initial
 condition (\ref{initialcondition}) at $b=0$.  Then, once an approximate
 solution was obtained, the $b$ dependence is restored, assuming similar
 $b$-dependences for both the solution and the initial
 conditions. Specifically, the following ansatz was used for the
 $b$-dependence of $\tn$:
\begin{equation} 
\label{Nb}  
\tn(r_\perp,x; b)\,=\, 
(1\,-\,e^{-\kappa(x,r_\perp)\, S(b)/S(0)})\,,  
\end{equation} 
where  
\begin{equation} \label{kappa} 
\kappa(x,r_\perp)\,=\,-\,\ln(1\,-\,\tn(r_\perp,x,b=0)).  
\end{equation} 
 
\section{Cross section for $\jpsi$ production} \label{sec:cross} 
The cross section for $\jpsi$ production is given by: 
\begin{equation}\label{eq:sigtot} 
\sigma(\gamma^*p\longrightarrow Vp) = \int d^2b 
\left|\int dz\, d^2\rt \Psi_{\gamma^*}(\rt,z,Q^2)\cA(\rt,x;b)\Psi_V(\rt,z)\right|^2, 
\end{equation} 
 where $\cA(\rt,x;b)$ is the imaginary part of the production amplitude in
 impact parameter space, and $\Psi_V$ and $\Psi_{\gamma^*}$ are,
 respectively, the wave functions of the $\jpsi$ and the virtual photon.
 Formally speaking $\Psi_V$ should depend on $b$. However, as the dominant
 contribition of the $\jpsi$ wave function comes from short distances, this
 dependence is neglected.  The evaluation of (\ref{eq:sigtot}) is done by
 first performing the polarization summation of the $\jpsi$-photon overlap
 function,$\Psi_V\times\Psi_{\gamma^*}$.  This overlap function can be
 derived in the $\rt$ representation using the spin structures of the vector
 meson wave function \cite{BM} and the well-known photon wave function
 \cite{WF}.

A detailed analysis of the overlap function has already been made in
 \cite{LMR}, where it was found that, in the momentum representaion, the
 overlap function for transverse (T) and longitudinal (L) polarized photon is
 given by \cite{LMR}:
\begin{equation}\label{overkt} 
\Psi_V(\kt,z)\,\times\,\Psi_{\gamma^*,T}(\kt,z)\propto 
\frac{2(z^2+(1-z)^2)\kt^2a^2+m_c^2(a^2-\kt^2)} 
{m_c\,(a^2+\kt^2)^3}\Psi_V(\kt,z)\,, 
\end{equation} 
\begin{equation}\label{overkl} 
\Psi_V(\kt,z)\,\times\,\Psi_{\gamma^*,L}(\kt,z)\propto 
\frac{2z(1-z)Q(a^2-\kt^2)}{(a^2+\kt^2)^3}\Psi_V(\kt,z)\,, 
\end{equation} 
 with $a^2=z(1-z)Q^2+m_c^2$. Here $\times$ denotes the polarization and
 helicity summation.  For a given $\jpsi$ spatial distribution, the
 transformation to configuration space is straightforward and is given in
 \eqs{overlapT}{overlapL} below.  Although many models exist for the spatial
 distributions of vector mesons (see for example,
 \cite{BM,Fermi,Kulzinger,Nemchik,Ivanov,Caldwell}), we choose to approximate
 $\Psi_V(\rt,z)$ to be $\Psi_V(\rt=0,z=\half)$. As in our previous
 publications \cite{GLMNpsi,GLMNpsi2}, we consider below a deviation from
 this approximation due to relativistic effects, produced by the Fermi motion
 of the bound quarks within the vector meson.

The effect of this motion, however, strongly depends on the charmed quark
 mass, $m_c$.  If one assumes, for example, that $m_c=M_{\psi}/2 \simeq
 1.55\gev$, then, by definition, there is no correction due to Fermi motion.
 On the other hand, in \cite{Fermi}, it has been assumed that $m_c \simeq
 1.50\gev$, and a suppression factor of the cross section of about $0.25$ was
 obtained, with almost no energy dependence.  Hence, as it stands, the
 contribution of this effect lies within a substantial range of uncertainty
 in which $m_c$ varies by no more than $0.05\gev$.

We, therefore, consider the effect of Fermi motion as an uncertainty of the
 wave function normalization, introducing it as an overall
 (energy-independent) suppression factor, $K_F$, which we use as a fitting
 parameter.  Using the above approximation we Fourier transform Eqs.\
 (\ref{overkt}) and (\ref{overkl}) and obtain:
\begin{eqnarray} 
\lefteqn{ 
\Psi_V(\rt=0,z=\half)\,\times\,\Psi_{\gamma^*,T}(\rt;Q^2)=  
\frac{K_F}{48\aem}\sqrt{\frac{3\Gamma_{ee}M_{\psi}}{\pi}}\,\times 
} \nonumber\\ 
& &  
\hspace{3cm}\left\{ 
\frac{a^2}{m_c}\, 
\left(\,\zeta\,K_1(\zeta) - \frac{\zeta^2}{4}K_2(\zeta) \,\right) 
+m_c 
\left(\frac{\zeta^2}{2}\,K_2(\zeta) - \zeta\,K_1(\zeta) \,\right) 
\right\} 
\label{overlapT} 
\end{eqnarray} 
\begin{equation}\label{overlapL} 
\Psi_V(\rt=0,z=\half)\,\times\,\Psi_{\gamma^*,L}(\rt;Q^2)=  
\frac{K_F}{48\aem}\sqrt{\frac{3\Gamma_{ee}M_{\psi}}{\pi}} 
\frac{Q}{2}\left(\frac{\zeta^2}{2}\,K_2(\zeta) - \zeta\,K_1(\zeta) \,\right). 
\end{equation} 
 where $\zeta=a\rt$, $K_i,i=1,2$ are the modified Bessel functions and
 $\Gamma_{ee}=5.26\,\mbox{KeV}$ is the leptonic width of the $\jpsi$.

Before presenting our results for the cross section (\ref{eq:sigtot}), we
 discuss the impact parameter dependence of the production amplitude,
 $\cA$. At first sight, $\cA(\rt,x;b)$ equals $\tn(\rt,x;b)$ where its
 $b$-dependence is related to the solution of \eq{eq:BK} at $b=0$ through the
 ansatz of \eq{Nb}.  The parameter $R$ in $S(b)$ [see \eq{dipS}] is related
 to the hadron size and has already been determined in \cite{NLE02} by
 fitting $\tn$ to the experimental data of $F_2$.  However, as stated in the
 introduction, a good fit to $F_2$ was obtained using a relatively small
 value of $R^2=4.5\gevs$, whereas we know that the the measured $\jpsi$
 differential slope, $B$, is consistent with a radius which is more than
 twice larger.

Hence, it turns out that $\cA$ differs from $\tn$.  The correct $b$-dependence
 of $\cA$ is given by the following expression: 
 
\begin{equation} \label{AB} 
\cA(\rt,x;b) = \int d^2b' \tn(\rt,x; b')\, S'(b-b')\, .
\end{equation} 
 
\begin{figure} 
\begin{center} 
\epsfig{file=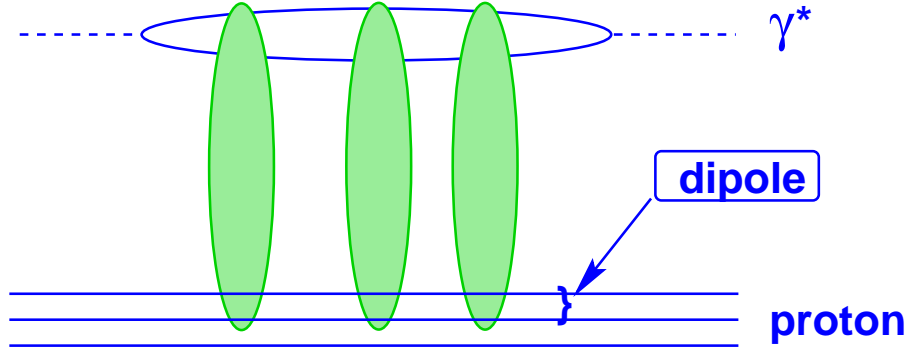,width=12cm,angle=0} 
\end{center} 
  \caption[]{\parbox[t]{0.80\textwidth}{\small 
Rescatterings for total DIS  cross section. 
}} 
\label{txsec} 
\end{figure}

To understand \eq{AB} we need to recall the procedure for calculating the 
 total cross section for the deep inelastic process which is shown in 
 \fig{txsec}.  If one uses the specific model in which the proton consists of 
 two colour dipoles (see Refs. \cite{DFK,BGLLM} for details), the total deep 
 inelastic cross section is given solely by the scattering amplitude of one 
 particular dipole present in the proton target.  More specifically, as there 
 is no momentum transfer which is involved in the process, the total cross 
 section does not depend on the probability of finding a second (spectator) 
 dipole in the target. 
 
\begin{figure} 
\begin{center} 
\epsfig{file=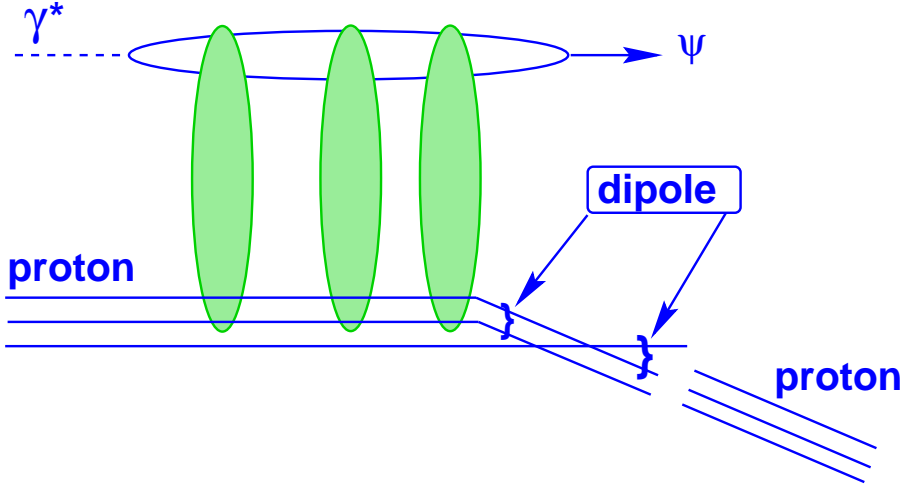,width=12cm,angle=0} 
\end{center} 
  \caption[]{\parbox[t]{0.80\textwidth}{\small 
Rescatterings for $J/\psi$ cross section at fixed $t = -q^2$ . 
}} 
\label{psisec} 
\end{figure}

On the other hand, in the process of $J/\psi$ production, the momentum
 transfer, $q$, is not zero (see \fig{psisec}).  In this process, for fixed
 momentum transfer, the $t=-q^2$ dependence of the $J/\psi$ production
 relates both to $\tn(\rt,x;q)$ (the amplitude for the photon to scatter off
 one dipole) and to the probability to find the second dipole having a
 momentum $q$ inside of the recoiled proton.  Denoting this probability by
 $S'(q)$, the production amplitude is thus proportional to
 $\tn(\rt,x;q)S'(q)$.  A product in momentum representation is equivalent to
 a convolution in the impact parameter, hence the form of \eq{AB}.
 
\begin{figure} 
\begin{center} 
\epsfig{file=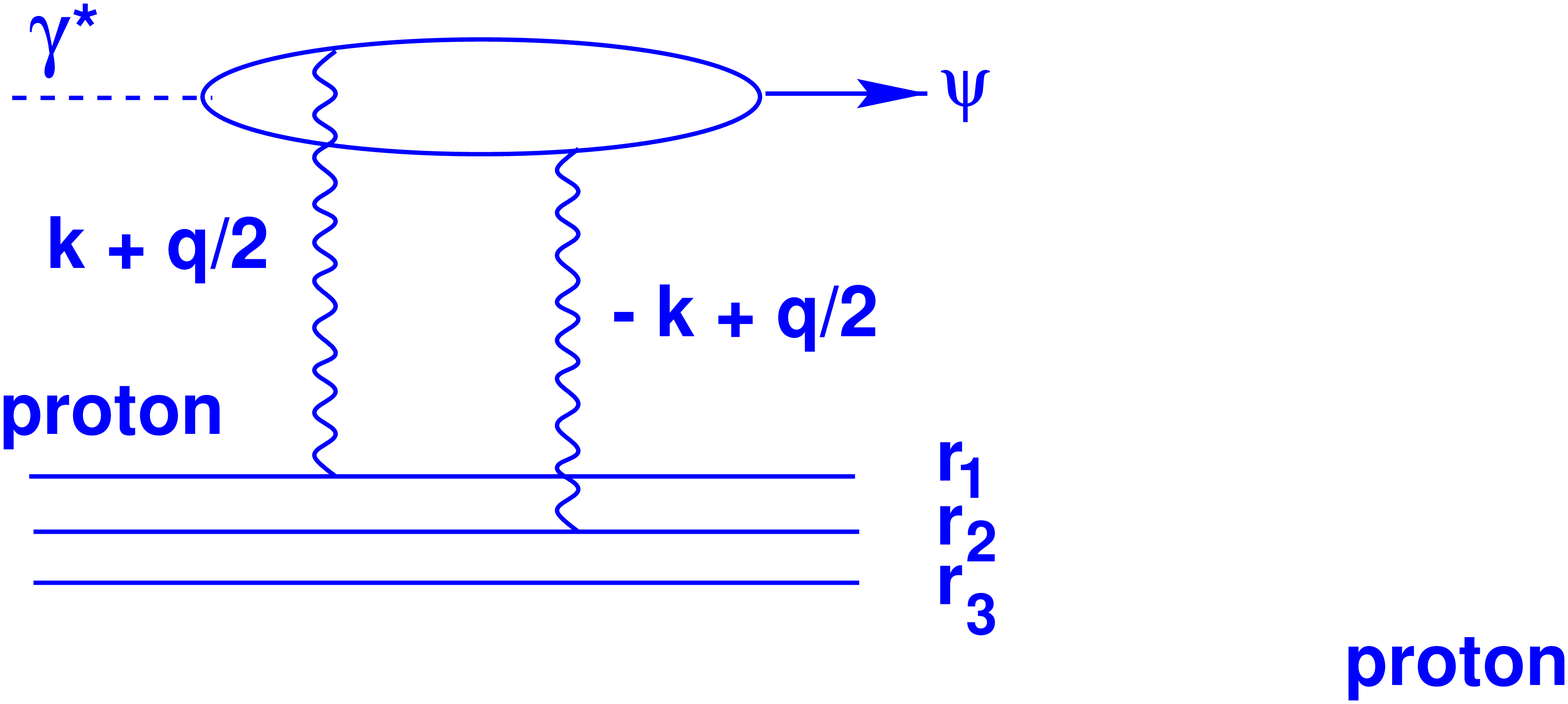,width=12cm,angle=0} 
\end{center} 
  \caption[]{\parbox[t]{0.80\textwidth}{\small 
The amplitude in the Born approximation
 for the  $J/\psi$ cross section at fixed $t = -q^2$. 
}} 
\label{bapsi} 
\end{figure}

Actually, we can estimate $S'(q)$ using the Born approximation for the 
 scatttering amplitude shown in the diagram of \fig{bapsi}. Assuming a  
 simple factorized form of the proton wave function, 
\begin{equation} 
\Psi_{\scrbox{proton}}(\rt,\rhot) 
\,=\,\phi(\rt)\,\phi(\rhot) 
\end{equation} 
where $\rhot = r_{1,\perp} - r_{2,\perp}$ and  
$\rt = r_{3,\perp} - (r_{1,\perp} + r_{2,\perp})/2$, 
we can easily calculate $S'(q)$ 
\begin{equation} \label{BAXSEC} 
S'(q)\,\,=\,\,\int d^2 \rt \,\left|\phi(\rt)\right|^2 \, 
e^{i\frac{1}{3}\vec{q}\cdot\vec{\rt}} 
\end{equation} 
Comparing \eq{BAXSEC} with the expression for the electromagnetic form 
factor for proton 
\begin{equation} \label{FFP} 
F(q)\,\,=\,\,\int d^2 \rt\,\left|\phi(\rt)\right|^2  
\,e^{i\frac{2}{3}\,\vec{q}\cdot\vec{\rt}} 
\end{equation} 
 one can see that $S'(b)$ can be described by \eq{dipS} with
 $R=R_{\scrbox{proton}}/2$, where $R_{\scrbox{proton}}$ is the
 electromagnetic radius of proton.
 
Alternatively, we can choose a different strategy and extract the typical
 size for the profile function $S'(b)$ using the experimental data.
 Generally, the differential slope is related to the average of the square of
 impact parameter, weighted by the the dipole-proton cross section.  The
 correct averaging procedure would be to integrate both over the impact
 parameter and over the dipole size, thereby reconstructing the energy
 dependence of $B$ (the shrinkage of the diffraction peak):
\begin{equation}\label{eq:2} 
B(x) = \half\langle b^2 \rangle =  
\frac 
{  \int d^2\rt b^2d^2b\, \tn(\rt,x,b)} 
{2 \int d^2\rt d^2b   \, \tn(\rt,x,b)} . 
\end{equation} 
 The value of $B$ as obtained from (\ref{eq:2}) is, of course, not universal 
 for all processes, and it directly depends on the choice of $S(b)$.  For the 
 purpose of calculating the effective radius to be used in $S'(b)$, we define 
 the deviation of $\half\langle b^2\rangle$ from the measured slope to be: 
\begin{equation}  
B' = B_{\scrbox{exp}} - \half\langle b^2\rangle, 
\end{equation} 
 where, $B_{\scrbox{exp}}$ is taken from experimental data (see, \eg, 
 \cite{ZEUSpsiEPS2001}). $S'(b)$ is then calculated 
 from \eq{dipS} with the substitution $R\longrightarrow R'=2B'$. 

\begin{figure} 
\begin{center} 
\epsfig{file=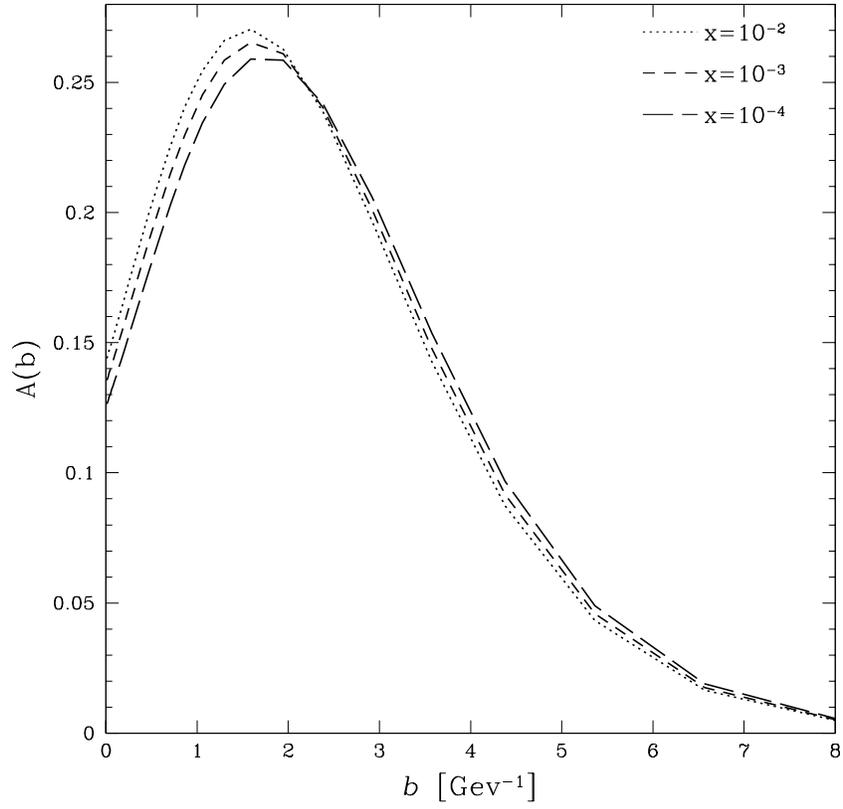,width=12cm,angle=0} 
\end{center} 
  \caption[]{\parbox[t]{0.80\textwidth}{\small 
The impact parameter dependence of the   $J/\psi$ production amplitude. 
}} 
\label{bdep} 
\end{figure}

The resulting $b$-dependence of $\cA(x;b)\equiv\int d^2\rt \cA(\rt,x;b)
 \Psi_{\gamma}(\rt)\Psi_V$ are shown in \fig{bdep}.  The convolution of
 $S'(b)$ with $\tn(\rt,x;b')$ is realized by a shift in the maximum of the
 scattering amplitude in the impact parameter space.  The decrease of the
 amplitude in the low $b$ region may be understood as a signature of
 shadowing corrections needed to insure unitarity.

Our calculations of $\sigma(\gamma^*p\longrightarrow\jpsi)$ were basically in
 accordance with \eq{eq:sigtot}, with modifications due to the contribution
 from the real part of the production amplitude \cite{EDEN} and the skewed
 (off diagonal) gluon distribution \cite{OFFDIAGONAL}.
The contribution of the real part is given by: 
\begin{equation} 
C_R^2 = (1\,+\rho^2)\,, 
\end{equation} 
where, 
\begin{equation}\label{er:realpart1} 
\rho = Re\cA/Im\cA\,=\,tg(\frac{\pi \lambda}{2}) 
\end{equation} 
and, in our approach, 
\begin{equation}\label{eq:realpart2} 
\lambda = \partial \ln (\tn)/\partial \ln (\frac{1}{x}). 
\end{equation} 
 Note that with this definition of $\lambda$, its value obeys the unitarity 
 constraint according to which, at a fixed value of $Q^2$, $\lambda$ is a 
 decreasing function of the energy with $\lambda\rightarrow 0$ as 
 $x\rightarrow 0$ \cite{NLE02}. 
 
The off diagonal contribution is given by: 
 \begin{equation}\label{eq:OD} C_G^2 = 
 \(\frac{2^{2\lambda+3}\,\Gamma(\lambda+\frac{5}{2})} 
 {\sqrt{\pi}\,\Gamma(\lambda+4)}\)^2. 
\end{equation} 
Formally, the definition (\ref{eq:realpart2}) of $\lambda$ is  
$\rt$-dependent.  In practice, the corrections due the $\rt$-dependence of 
$\lambda$ are rather small.  Thus, $\lambda$ 
is computed at fixed scale $\rt^2=4/(Q^2+M^2_{\jpsi})$. 

\begin{figure} 
\begin{center} 
\epsfig{file=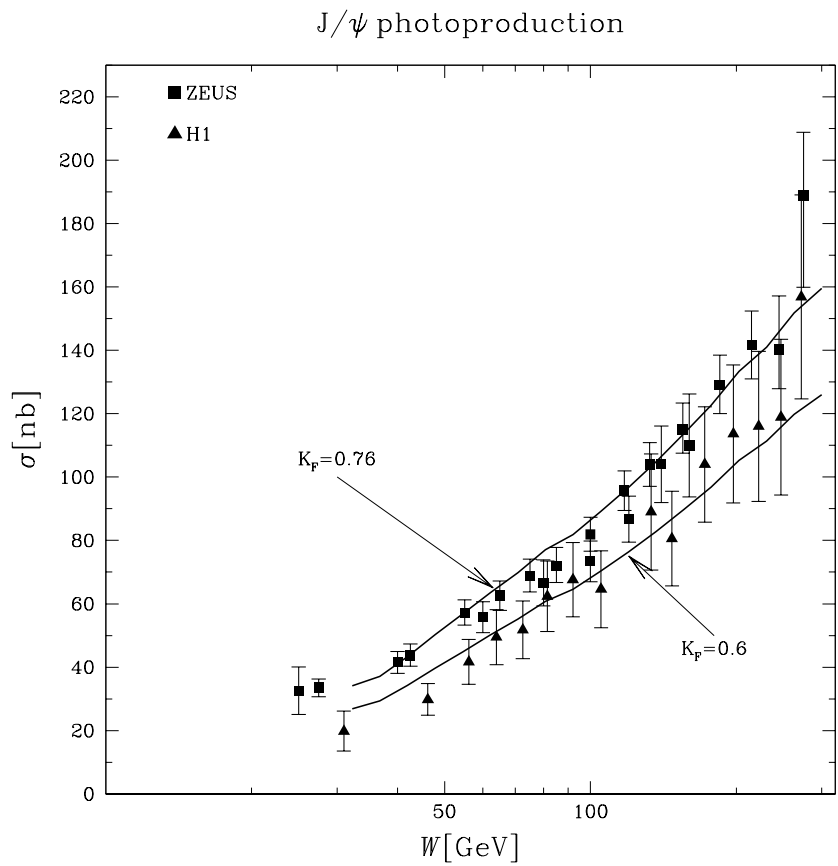,width=12cm,angle=0} 
\end{center} 
  \caption[]{\parbox[t]{0.80\textwidth}{\small 
 The ZEUS and H1 $J/\psi$ photoproduction data compared with our calculations 
using NLE.  
}} 
\label{fig:photo} 
\end{figure}

The results of our calculations of \eq{eq:sigtot}, given the above
 modifications, are shown in \fig{fig:photo} for $\jpsi$ photoproduction,
 where we define $x=M_{\jpsi}^2/W^2$. We found that the H1 data
 \cite{H1psi2000} are compatible with $K_F=0.6$ ($\chisquare=0.5$) while the
 ZEUS data require $K_F=0.76$ ($\chisquare=0.6$).  As stated, the factor
 $K_F$ is the deviation of our approximation for the $\jpsi$ static potential
 wave function from a more realistic model, in which the spatial distribution
 of $\jpsi$ depends on $\rt$ and $z$.  Our results are consistent with
 \cite{Caldwell}, where a comparison between a realsitic and static $\jpsi$
 wave functions was made.

\begin{figure} 
\begin{center} 
\epsfig{file=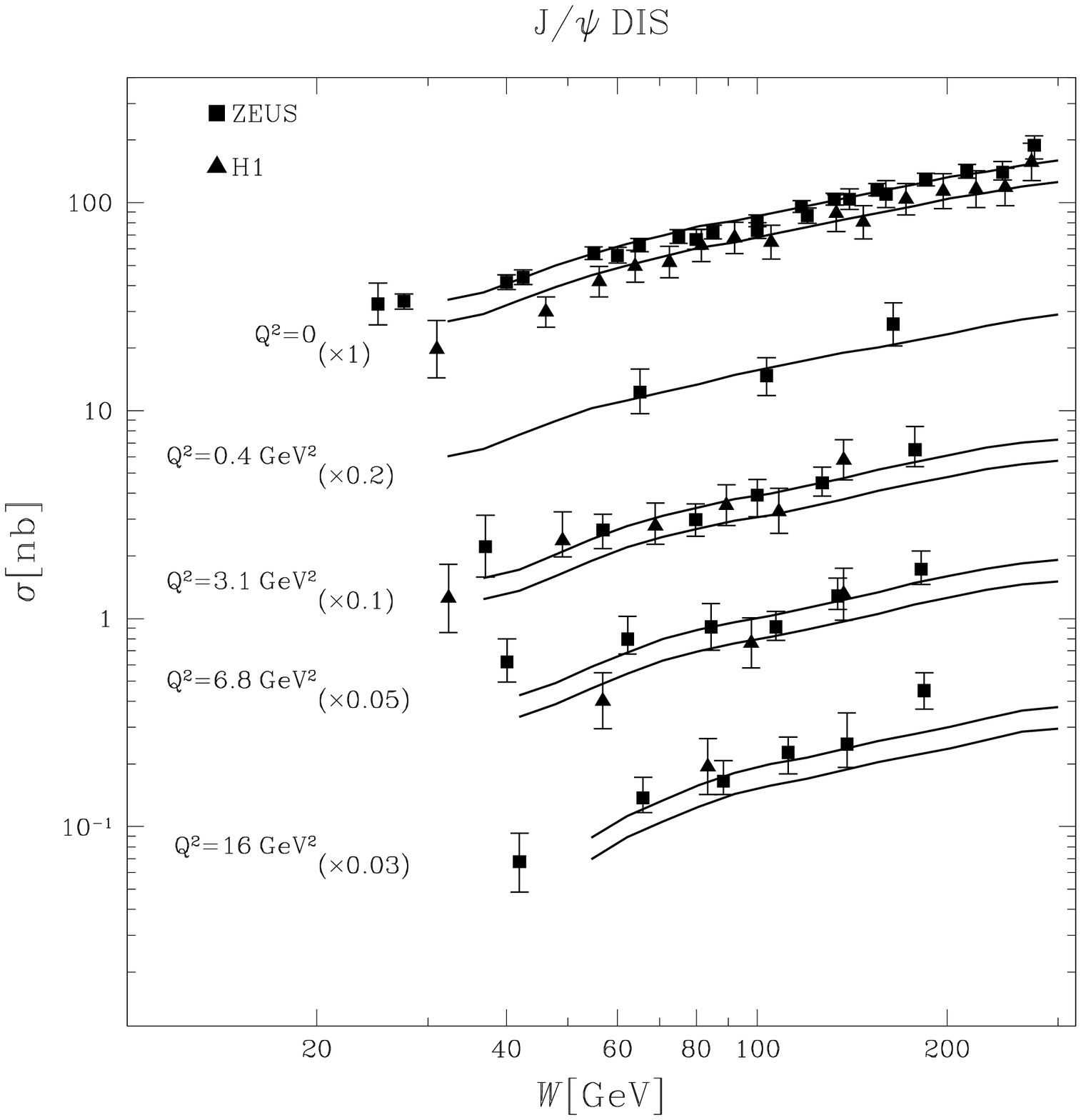,width=12cm,angle=0} 
\end{center} 
  \caption[]{\parbox[t]{0.80\textwidth}{\small 
$J/\psi$ DIS and photoproduction, data and our calculations using NLE.  The 
upper (lower) curves at each $Q^2$ bin correspond to different  $K_F$ 
values, which were fitted to ZEUS (H1) photoproduction data. 
}} 
\label{fig:DIS} 
\end{figure}

The values of $K_F$ which were obtained for $Q^2=0$, were used to predict the
 cross section for $Q^2>0$.  In \fig{fig:DIS} we compare the $Q^2>0$
 predictions to the preliminary data which were read off plots in
 \cite{ZEUSpsiDIS2002}.  Within the experimental errors, our reproduction of
 the data is good, except for data at $Q^2=16\gevs$ where we underestimate
 the H1 point ($W\simeq 80\gev$) and one of the ZEUS points ($W\simeq
 180\gev$).  In spite of this, the overall $\chisquare$ for photoproduction
 and DIS data (including these two problematic points) is 0.73 for ZEUS and
 0.84 for H1.
 
 
 \section{Predictions for Heavy Nuclei} \label{sec:nuc}

In this section we assess the unitarity effects in $\jpsi$ coherent
 production in photo and DIS on a nuclear target. In such processes,
 unitarity effects are more pronounced than in $e-p$ collisions. Thus, with
 nuclear targets one can access the region of hdQCD at values of $x$ which
 are larger than those characterizing this region at HERA experiments.

The production of vector mesons on a nuclear target can be calculated in a
 straightforward manner using methods similar to the above.  In principle,
 (\ref{eq:BK}) can be solved separately for each atomic number $A$, with the
 same form of initial conditions.  To this end we choose to present our
 predictions using the Glauber approach, and for the time being to postpone
 the solution of (\ref{eq:BK}) for nuclei.

In the Glauber approach, the cross-section for the $\jpsi$ production,
 including contributions for the percolation through a nucleus of both a
 $\ket{q\bar{q}}$ system and a $\ket{q\bar{q}g}$ system, is given by the
 following expression:
\newcommand{\GLAUBERI}{\left(1-e^{-\half\Omega_q}\right)}
\newcommand{\GLAUBERII}{
\frac{C_F}{\pi^2}\as\rt^2\int\frac{dx}{x}
\int_{\rt'>\rt}\frac{d^2\rt'}{\rt'^4}\left(1-e^{-\half\Omega_g}\right)
}
\begin{eqnarray}\label{eq:Glauber}
\sigma(\gamma^*A\longrightarrow \jpsi A) = \int d^2b 
\left|
\int dz\, d^2\rt \Psi_{\gamma^*}(\rt,z,Q^2)\Psi_{\jpsi}(\rt,z)\,\times
\right. \hspace{2.5cm} 
\nonumber \\
\left.
\left[\GLAUBERI + \GLAUBERII \right]
\right|^2,
\end{eqnarray}
where the opacities $\Omega_q$ and $\Omega_g$ are defined as:
\begin{eqnarray}\label{omegaq}
\Omega_q(b,\rt,x) &=& 
\frac{\pi^2}{3}\rt^2\as\(\frac{4}{\rt^2}\)xG\(x,\frac{4}{\rt^2}\)S_A(b)\,,
\\
\Omega_g(b,\rt,x)&=&\frac{9}{4}\Omega_q(b,\rt,x), 
\end{eqnarray}
 $S_A(b)$ is the number of nucleons in a nucleus interacting with the
 incoming dipole and $xG$ is the gluon density obtained from linear DGLAP
 evolution (we used CTEQ parameterization \cite{CTEQ} for this particular
 computation).

We are cognisant of the fact that when dealing with heavy nuclei
 interactions, neither a dipole nor a Gaussian are adequate for $S_A(b)$.
 When the probing hadronic system percolates through a heavy nucleus, it
 experiences multiple rescatterings over a substantially long region of the
 impact parameter space. On the other hand, once the system leaves the heavy
 nuclei, the number of strong interactions decreases rapidly.

Our approach is to use the Wood-Saxon parameterization \cite{Enge} for the 
impact parameter dependence, $S_A(b)$: 
\begin{equation}\label{eq:WS1} 
S_A(b) = \rho\int\,\frac{d r_{||}}{1+e^{\frac{r-R_A}{h}}} 
\end{equation} 
 where $r_{||}$ is the longitudinal distance from the target, $h$ and $R_A$ are 
 parameters which are taken from experimental tables \cite{NUCTAB}, 
 $r=\sqrt{r_{||}^2+b^2}$ and the normalization factor $\rho$ is defined in the 
 following relation: 
\begin{equation}\label{eq:WS2} 
\rho\int\,\frac{dr_{||}\,d^2b}{1+e^{\frac{r-R_A}{h}}} = A\,. 
\end{equation} 
 Note that $R_A$ is associated with the radius of a nucleus with atomic
 number $A$.

\begin{figure} 
 \begin{center} 
\epsfig{file=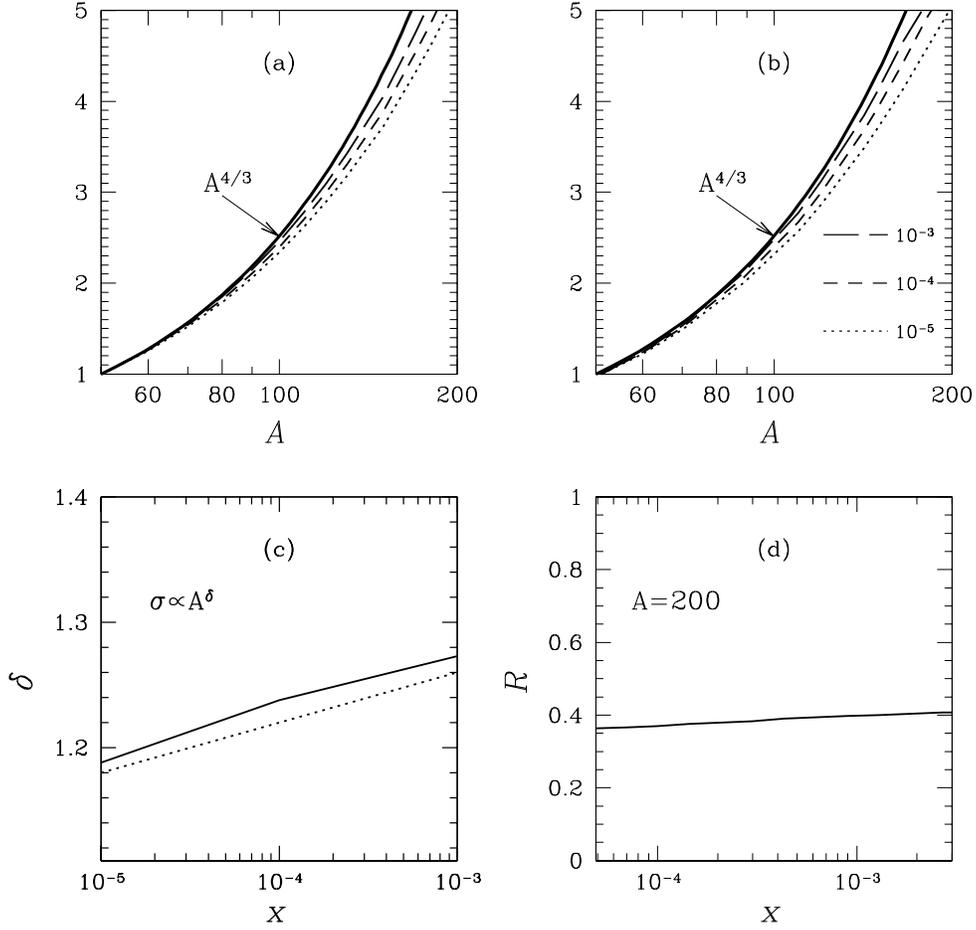,width=14cm,angle=0}
\end{center} 
  \caption[]{\parbox[t]{0.80\textwidth}{\small 
$J/\psi$ photoproduction on a nuclear target: (a)-(b) show the $A$-dependence
 of the cross section for different values of $x$, normalized to unity at
 $A=50$ [(a) shows the contribution due to the interactions of
 $\ket{q\bar{q}}$ and $\ket{q\bar{q}g}$ with the nucleus, and (b) shows the
 contributions of all interactions]; (c) shows the values of $\delta$, as a
 function of $x$, where the lower curve corresponds to plot (a) and the upper
 curve corresponds to plot (b); and (d) shows the normalization ratio between
 plot (a) and plot (b) as  explained in the text.
}}
\label{fig:psinuc} 
\end{figure}

To determine the variance between nuclei targets and a nucleon target, we
 need to examine the $A$ dependence of the integrated cross section, by
 comparing the calculated cross section to a power behaviour of
 $\sigma\propto A^\delta$.  The maximal value of the exponent,
 $\delta_{max}$, depends on the profile function $S_A(b)$, and can be found
 from the following integral:
\begin{equation}
A^{\delta_{max}} = \int d^2b\left|S_A(b)\right|^2
\end{equation}
 For an exponential profile, $\delta_{max}=\frac{4}{3}$ and for a Wood-Saxon
 profile of (\ref{eq:WS1}), $\delta_{max}\approx 1.43$. A deviation from the
 maximal value of $\delta$ would be a clear signature of saturation of the
 growth of the cross section.  \fig{fig:psinuc}a shows the cross section as a
 function of $A$ with an arbitrary normalization, for $x=10^{-5}, 10^{-4}$
 and $10^{-3}$.  To illustrate the saturation effect, we compare the curves
 to a power behavior of $\sigma\propto A^{4/3}$.  Unitarity conserving
 effects are already appreciable at $x=10^{-3}$.

\eq{eq:Glauber} takes into account the interaction of the quark-antiquark
 pair and the fastest gluon. On the other hand, we can improve our Glauber
 approach by using the solution of the non-linear equation, since the
 solution to the non-linear equation takes into account all possible
 interactions of partons with the nucleon target. As stated, $R^2_p$, the
 effective radius of nucleon turns out,in our approach, to be very
 small. With such a small radius, the typical parameter that governs the
 dipole interaction with the nuclear target is also very small, namely,
\begin{equation} \label{TYPA}
R^2_p\,S_A(0)\,\,\leq\,\,1\,.
\end{equation}
 In the standard approach, however, $R^2_p\,S_A(0)$ was considered to be
 large. We thus have to take into account all rescatterings inside
 the proton in the first stage of our approach and treat the interactions
 with different nucleons in a nucleus using the  Glauber formula.

Specifically, the cross section can be written in the form: 
\begin{eqnarray}\label{eq:MGlauber}
\lefteqn{\sigma(\gamma^*A\longrightarrow \jpsi A) =} \nonumber \\
& & \int d^2b
\left|
\int dz\, d^2\rt\,\Psi_{\gamma^*}(\rt,z,Q^2)\Psi_{\jpsi}(\rt,z)\,
\left(1 - e^{- \frac{1}{2}\sigma(dipole-nucleon) \,S_A(b)}\right) 
\right|^2
\end{eqnarray}
where $\sigma(dipole-nucleon)$ is equal to 
$2\,\int\,d^2\,b\,N(r^2,x;b)$ where $N$ is dipole-nucleon amplitude.

The following figures illustrate the effect of the inclusion of all
 rescatterings.  \fig{fig:psinuc}b shows the cross section of
 \eq{eq:MGlauber} as a function of $A$ with an arbitrary normalization, for
 $x=10^{-5}, 10^{-4}$ and $10^{-3}$. 

In \fig{fig:psinuc}c we display the characteristic exponent, $\delta$, as a
 function of $x$, for the calculated cross-sections of Eqs.\
 (\ref{eq:Glauber}) and (\ref{eq:MGlauber}).  The upper curve in
 \fig{fig:psinuc}c corresponds to the numerical calculation of
 \eq{eq:MGlauber} and the lower curve corresponds to \eq{eq:Glauber}. In
 terms of $\delta$, the differences between the two approaches are
 negligible.  On the other hand, as further detailed below, there is a
 considerable normalization difference.

To investigate the normalization difference, we define
 the following ratio:
 \begin{equation}\label{eq:R}
 {\cal R}=\frac{\eq{eq:MGlauber}}{\eq{eq:Glauber}}
 \end{equation}
 We have calculated ${\cal R}$ for $\jpsi$ photoproduction on a Gold target
 ($A\simeq 200$).  The results, as a function of the energy, for several
 values of $Q^2$, are shown in \fig{fig:psinuc}d.  Our calculations show that
 the full calculation is suppressed by a factor of about 0.4 with respect to
 the calculation in which only a $\ket{q\bar{q}}$ and a $\ket{q\bar{q}g }$
 systems are considered to interact with the nucleus.

\begin{figure} 
 \begin{center} 
\epsfig{file=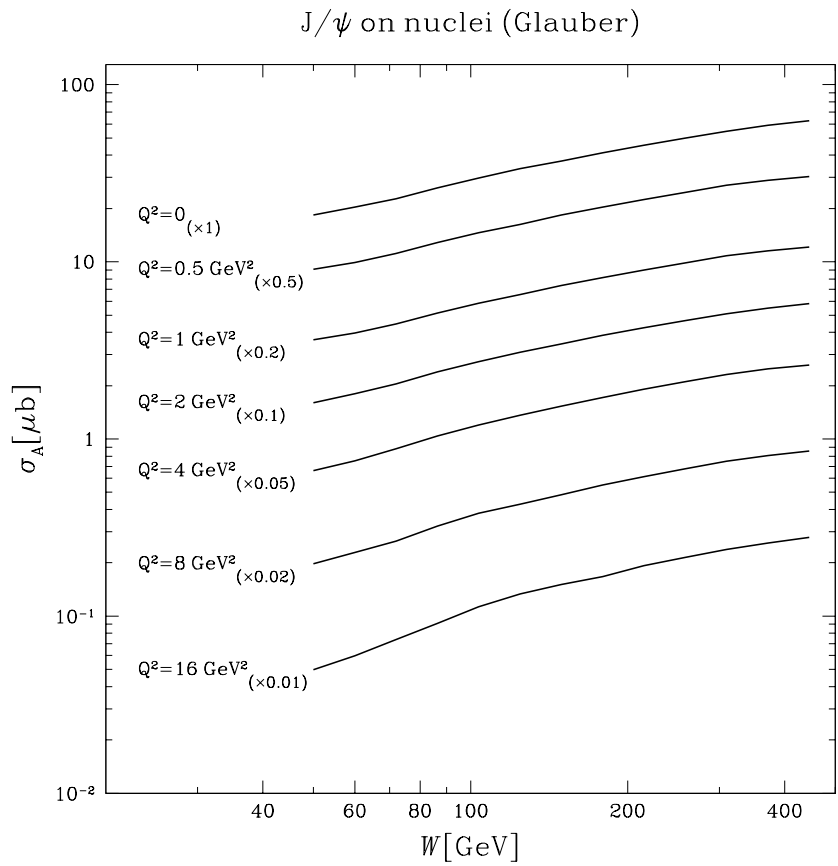,width=12cm,angle=0}
\end{center} 
  \caption[]{\parbox[t]{0.80\textwidth}{\small 
Predictions fof $J/\psi$ photo- and DIS production on a Gold target, using 
the
Glauber approach of \eq{eq:MGlauber} for 
different values of $Q^2$ as a function of $W$.  
}} 
\label{fig:psinucdis} 
\end{figure}

Our prediction for the energy dependence for $\jpsi$ production on a Gold
 target, in the calculation which includes all parton interactions is shown
 in \fig{fig:psinucdis} for several values of $Q^2$.  In this figure, we have
 used the wave function normalization which we have extracted from the data
 on proton target.  Due to the uncertainty of this normalization, and to the
 0.4 ratio of \eq{eq:R},our rough estimation of the relative errors of the
 curves are of the order of 15-20\percent.

 \section{Summary} \label{sec:summary}
 
We have extended our recent investigation of the approximate solution of the
 nonlinear evolution equation to the case of $\jpsi$ photo- and DIS
 production. We have convoluted our previous ansatz for the $b$-dependence of
 the amplitude with an additional profile which is extracted from the
 electromagnetic form factor.  The resulting impact parameter dependence of
 the amplitude exhibits a decrease near $b=0$. We believe that this decrease
 is due to the deviation from the linear evolution equations, and is a
 signature for the onset of unitarity taming effects.
 
We have used the solution to the NLE and obtained a satisfactory
 reproduction of both the photo- and the DIS production data.  Our free 
 parameter was the uncertainty in the $\jpsi$ wave function normalization due 
 to Fermi motion of the $c\bar{c}$ pair.  In addition, we took into account 
 modifications of the cross section due to the real part of the amplitude and 
 the existence of skewed gluons within the nucleon. 

We used the Glauber approach to calculate the cross section for $\jpsi$
 production on a nuclear target.  Our predictions demonstrate that such
 processes are important for the understanding of unitarity taming
 effects. We have found that these effects start to dominate for
 $x\le10^{-3}$. Hence, the predictions presented here may be verified with
 the future eRHIC experiments.

It is important to note that, apart from the uncertainities due to vector
 meson wave function, the calculations on nuclei do not contain any fitting
 parameters. More specifically the uncertainity due to $b$ dependence of the
 solution for nuclei is quite small.  In other words, the initial conditions
 for the BK equation \cite{BA,KO} are determined solely by the proton DIS
 data.
 
\section*{Acknowledgments} 

{\bf This paper is dedicated to our friend and colleague Jan Kwiecinski
 on the occasion of his sixty-fifth birthday, may he continue to be active
 and contribute for many years to come.} 

Two of us (E.G. and E.L.) thank the DESY Theory Division for their
 hospitality.  This research was supported in part by GIF grant $\#$
 I-620-22.14/1999 and by Israeli Science Foundation, founded by the Israeli
 Academy of Science and Humanities.

 
\end{document}